\input harvmac.tex
%
\figno=0
\def\fig#1#2#3{
\par\begingroup\parindent=0pt\leftskip=1cm\rightskip=1cm\parindent=0pt
\baselineskip=11pt \global\advance\figno by 1 \midinsert
\epsfxsize=#3 \centerline{\epsfbox{#2}} \vskip 12pt {\bf Fig.
\the\figno:} #1\par
\endinsert\endgroup\par
}
\def\figlabel#1{\xdef#1{\the\figno}}
\def\encadremath#1{\vbox{\hrule\hbox{\vrule\kern8pt\vbox{\kern8pt
\hbox{$\displaystyle #1$}\kern8pt} \kern8pt\vrule}\hrule}}

\overfullrule=0pt

%
\def\cqg#1#2#3{{\it Class. Quantum Grav.} {\bf #1} (#2) #3}
\def\np#1#2#3{{\it Nucl. Phys.} {\bf B#1} (#2) #3}
\def\pl#1#2#3{{\it Phys. Lett. }{\bf B#1} (#2) #3}

\def\cmp#1#2#3{{\it Comm. Math. Phys.} {\bf #1} (#2) #3}

\font\zfont = cmss10 
 
\def\bigone{\hbox{1\kern -.23em {\rm l}}}
\def\ZZ{\hbox{\zfont Z\kern-.4emZ}}

\def\a{\alpha}
\def\b{\beta}
\def\g{\gamma}
\def\d{\delta}
\def\e{\epsilon}

\def\m{\mu}
\def\n{\nu}
\def\x{\xi}
\def\r{\rho}

\def\ps{\psi}
\def\G{\Gamma}
\def\D{\Delta}

\def\O{\Omega}
\def\o{\over}

\Title{ {\vbox{ \rightline{\hbox{hepth/0308014}}
\rightline{\hbox{UMD-PP-03-059}} }}}
{\vbox{\hbox{\centerline{PP-Waves, ${\cal M}$-Theory and Fluxes}}
}}
\smallskip
\centerline{{\bf Katrin Becker}\footnote{$^1$} {\tt
katrin@physics.utah.edu}}
\smallskip
\centerline{\it Department of Physics, University of Utah, Salt
Lake City, UT 84112-0830}

\bigskip \centerline{{\bf Melanie Becker and Ram
Sriharsha}\footnote{$^2$} {\tt melanieb@physics.umd.edu and
harsha@glue.umd.edu} } \centerline{\it Department of Physics,
University of Maryland} \centerline{\it College Park, MD
20742-4111}

\bigskip
\bigskip
\baselineskip 18pt
\bigskip
\noindent

We study a new type of warped compactifications of ${\cal
M}$-theory on eight manifolds for which nowhere vanishing
covariantly constant spinors of indefinite chirality on the
internal manifold can be found. We derive the constraints on the
fluxes and the warp factor following from supersymmetry and the
equations of motion. We show, that the lift of Type IIB PP-waves
to ${\cal M}$-theory is a special type of solution of this general
class of models. We comment on the relation between the Type IIB
version of such compactifications as a dual description of the
Polchinski-Strassler solution describing a four-dimensional
confining gauge theory.

\Date{July, 2003}

\newsec{Introduction and Summary}
Although warped compactifications have been known in string theory
for almost twenty years (see e.g. \ref\wnw{B.~De Wit, H.~Nicolai
and N.~P.~Warner, ``The embedding of Gauged $N=8$ Supergravity
into $d=11$ Supergravity, \np {255} {29} {1985}.},
\ref\str{A.~Strominger, ``Superstrings with Torsion'', \np {274}
{253} {1986}.} and \ref\wsd{B. de Wit, D.~J.~Smit and N.~D.~Hari
Dass, ``Residual Supersymmetry of Compactified $D=10$
Supergravity'', \np {283} {1987} {165}.}) \lref\HULL{C.~M.~Hull,
{\it ``Superstring Compactifications with Torsion and Space-Time
Supersymmetry,''} In Turin 1985, Proceedings, Superunification and
Extra Dimensions, 347-375, 29p; {\it ``Sigma Model Beta Functions
and String Compactifications,''} Nucl.\ Phys.\ B {\bf 267}, 266
(1986); {\it ``Compactifications of the Heterotic Superstring,''}
Phys.\ Lett.\ B {\bf 178}, 357 (1986); {\it ``Lectures on
Nonlinear Sigma Models and Strings,''} Lectures given at Super
Field Theories Workshop, Vancouver, Canada, Jul 25 - Aug 6, 1986.}
they have become recently an active area of research as it has
been realized, that these compactifications are excellent
candidates to solve one longstanding problem in string theory, the
moduli space problem. Thus they are of great interest from the
phenomenological point of view, which is the reason of why these
compactifications were considered in the context of the heterotic
string in \ref\beckerD{ K.~Becker and K.~Dasgupta, { ``Heterotic
Strings with Torsion,''} JHEP {\bf 0211}, 006 (2002),
hep-th/0209077.}, \ref\bbdp{K.~Becker, M.~Becker, K.~Dasgupta and
S.~Prokushkin, {``Properties of Heterotic Vacua from
Superpotentials,''} hep-th/0304001.} and \ref\bbdg{K.~Becker,
M.~Becker, K.~Dasgupta and P.~S.~Green, { ``Compactifications of
Heterotic Theory on non-Kaehler Complex Manifolds.  I,''} JHEP
{\bf 0304}, 007 (2003), hep-th/0301161.}. For a recent discussion
on the subject of counting flux vacua and the moduli space problem
in string theory see \ref\ad{S.~Ashok and M.~R.~Douglas{
``Counting Flux Vacua''}, hep-th/0307049.}. Since the internal
manifold is in this case non-K\"ahler and has torsion these models
are also attractive from the mathematical point of view \ref\GP{
E.~Goldstein and S.~Prokushkin, {``Geometric Model for Complex
non-Kaehler Manifolds with SU(3) Structure,''} hep-th/0212307.},
{\bbdg} and \ref\carluest{ G.~L.~Cardoso, G.~Curio, G.~Dall'Agata,
D.~Lust, P.~Manousselis and G.~Zoupanos, {``Non-K\"ahler String
Backgrounds and their Five Torsion Classes,''} hep-th/0211118.}.
This is only one of the reasons, of why warped compactifications
are fascinating, as there are several connections to other
scenarios, that are as surprising and interesting as the previous
one.

First, as we shall see in this paper, there is a connection
between a generalization of the warped compactifications of ${\cal
M}$-theory on eight-manifolds considered in \ref\bb{K.~Becker and
M.~Becker, ``${\cal M}$-Theory on Eight Manifolds'', \np {477}
{1996} {155}, hep-th/9605053.} and the ${\cal M}$-theory lift of
PP-waves of the Type IIB theory considered in \ref\cv{M.~Cvetic,
H.~Lu and C.~N.~Pope, ``Penrose Limits, PP-Waves and Deformed
M2-Branes'', hep-th/0203082.}. We will show, that the construction
of {\cv} can be viewed as a ``compactification'' of ${\cal
M}$-theory on an eight manifold, where the internal manifold is
non-compact. Indeed, the models considered in {\cv} can be thought
of as ${\cal M}$-theory compactifications, where the internal
manifold admits only self-dual fluxes, differing from {\bb} by the
fact, that the internal spinors are non-chiral. To show the
existence of this more general class of models and to derive the
generic equations describing them will be the main point of the
present paper.

>From a different perspective, warped compactifications play also
an important role in the description of confining supersymmetric
gauge theories and ultimately in the description of QCD. This is
because there is a close relation between warped compactifications
and Ramond-Ramond backgrounds in string theory. Confining gauge
theories can be realized, for example, as perturbations by
three-form flux of Type IIB string theory on $AdS_5 \times S_5$.
It was shown by Polchinski and Strassler \ref\ps{J.~Polchinski and
M.~J.~Strassler, ``The String Dual of a Confining Four-Dimensional
Gauge Theory'', hep-th/0003136.}, that the three-form flux of the
supergravity theory corresponds to a perturbation of the ${\cal
N}=4$ gauge theory by mass terms and the resulting gauge theory
has ${\cal N}=1$ supersymmetry {\ps}.

Non-perturbatively, ${\cal N}=1$ supersymmetric gauge theories can
be realized by placing D3-branes at conical singularities of a
Ricci-flat six-dimensional cone, whose base manifold is a
five-dimensional Einstein space $X_5$. On the supergravity side
one considers the Type IIB theory on $AdS_5 \times X_5$ and this
is dual to the world-volume theory of the D3-branes at the
singularity. In case that one considers D3-branes on the conifold
\ref\kw{I.~R.~Klebanov and E.~Witten, ``Superconformal Field
Theory on Threebranes at a Calabi-Yau Singularity'', \np {536}
{1998} {199}, hep-th/9807080.}, for example, one would obtain on
the worldvolume of the D3-branes a gauge theory with $SU(N) \times
SU(N)$ gauge group. Besides considering D3-branes it is also
possible to consider D5-branes wrapped on collapsed two-cycles at
the singularity \ref\kt{I.~R.~Klebanov and A.~A.~Tseytlin,
``Gravity Duals of Supersymmetric $SU(N) \times SU(N+M)$ Gauge
Theories'', \np {578} {2000} {123}.}. This has the effect, that
the D3-brane charge eventually becomes negative and the
supergravity metric becomes singular. It was argued by Klebanov
and Strassler \ref\ks{I.~R.~Klebanov and M.~J.~Strassler,
``Supergravity and a Confining Gauge Theory: Duality Cascades and
$\chi$SB-Resolution of Naked Singularities'', JHEP {\bf 0008}
(052) 2000, hep-th/0007191.}, that this naked singularity of the
metric gets resolved in terms of a warped deformed conifold, which
is completely non-singular. It was realized later on in
\ref\gp{M.~Gra\~na and J.~Polchinski, ``Supersymmetric 3-Form Flux
and Perturbations of $AdS(5)$'', hep-th/0009211.} and
\ref\gub{S.~S.~Gubser, ``Supersymmetry and F-Theory Realization of
the Deformed Conifold with 3-Form Flux'', hep-th/0010010. }, that
the Klebanov-Strassler model can be obtained as a special case of
the solutions derived in {\bb} and \ref\svw{S.~Sethi, C.~Vafa and
E.~Witten, ``Contraints on Low Dimensional String
compactification'', \np {480} {1996} {213}, hep-th/9606122.},
describing compactifications of ${\cal M}$-theory on
eight-manifolds. This is interesting and one may wonder, if there
is a similar connection between the models considered in {\bb} or
a corresponding generalization thereof and the
Polchinski-Strassler model. This would be useful to derive the
exact solution of the model considered in {\ps}. In this paper we
will take one step in this direction. We shall consider
compactification of ${\cal M}$-theory on eight-manifolds for which
we can define two covariantly constant spinors on the internal
manifold of non-definite chirality. This is one generalization of
{\bb}, that is needed in order to make contact with {\ps}.

This paper is organized as follows. In section 2.1 we give a short
reminder on compactifications of ${\cal M}$-theory on
eight-manifolds, where two covariantly constant spinors of
definite chirality on the internal manifold can be found. These
manifolds are conformally Calabi-Yau. In section 2.2 we discuss
the generalization of this type of compactifications to the case,
where the spinors on the internal manifold no longer have a
definite chirality. In section 3 we discuss a particular example
of the models considered herein, namely the ${\cal M}$-theory lift
of the PP-wave solution of the Type IIB theory considered in {\cv}
and show, that it obeys the constraints, that we derived in the
previous section. In section 4 we give our conclusions and outlook
and comment on the relation between the present work as a dual
description of the Polchinski-Strassler model describing a
four-dimensional confining gauge theory. In an appendix we collect
some of the relevant formulas.

{\bf Note Added}: While this paper was written there appeared two
interesting papers, which have some overlap with the discussion
presented herein \ref\ms{D.~Martelli and J.~Sparks,
``G-Structures, Fluxes and Calibrations in M-theory'',
hep-th/0306225.} and \ref\fg{A.~R.~Frey and M.~Gra\~na, ``Type IIB
Solutions with Interpolating Supersymmetries'', hep-th/0307142.}.
The work of {\ms} discusses the most general supersymmetric
geometries arising in M-Theory compactifications on eight
manifolds, when the chirality assumption on the internal spinors
is removed.

\newsec{${\cal M}$-theory Compactifications on Eight Manifolds}
In this section we would like to consider compactifications of
${\cal M}$-theory on eight-manifolds and derive the constraints on
the fluxes and the warp factor, that follow from supersymmetry and
the equations of motion.

\subsec{Reminder: Chiral Spinors on the Internal Space}

Let us start with a short reminder about compactifications with
fluxes for which two spinors on the internal space, that are
chiral can be found {\bb}, {\svw}. This will be useful to
introduce our notation and will be helpful in order to compare
with the non-chiral case discussed afterwards. The bosonic part of
the action of the eleven-dimensional supergravity limit of ${\cal
M}$-theory is given by
 \ref\cjs{E.~J.~Cremmer, B.~Julia and J.~Scherk,
``Supergravity Theory in 11 Dimensions'', \pl {5} {1978} {409}.}
\eqn\ai{ {\cal S}_{11}={1\o 2} \int d^{11} x \sqrt{-{\hat g}
}\left[ {\hat R} -{1\o 2} \hat F \wedge * \hat F -{1\o 6 } \hat C
\wedge \hat F \wedge \hat F \right]   ,  }
where ${\hat g}_{MN} $ is the space-time metric (the hat denotes
eleven-dimensional quantities) and $\hat C$ is a three-form with
field strength $\hat F=d\hat C $. We have set the gravitational
constant equal to one. The complete action is invariant under
local supersymmetry transformations
\eqn\aii{ \eqalign{ & \d {\hat e_M}^A=i{\bar \eta} {\hat \G}^A
\psi_M,  \cr & \d \hat C_{MNP}=3 i{\bar \eta} {\hat \G}_{[MN} \psi
_{P]}, \cr & \d \psi_M ={\hat \nabla} _M \eta - {1\o 288} \left(
{{\hat \G}_M}^{\,\,\,\,\, PQRS}-8 {\hat \d}_M^P {\hat \G}^{QRS}
\right) \hat F_{PQRS} \eta,\cr} }
where ${\hat e_M}^A$ is the vielbein, $\psi_M$ is the gravitino,
$\eta$ is an eleven-dimensional anticommuting Majorana spinor and
${\hat\nabla}_M$ denotes the covariant derivative involving the
Christoffel connection as usual. Further notations and conventions
will be given in the appendix.

The field strength obeys the Bianchi identity
\eqn\aaiaai{d\hat F=0,}
or in components $\partial_{[M} \hat F_{PQRS]}=0$. This equation
is metric independent. The field equation for $\hat F$ is
\eqn\aixxiaa{ d * \hat F=-{1\o 2} \hat F^2,  }
or in components after dualizing
\eqn\aiv{\hat E^{-1} {\partial}_M ({\hat E} {\hat F}^{MNPQ})-{1 \o
1152} {\hat \e}^{NPQRSTUVWXY}
 \hat F_{RSTU} \hat F_{VWXY}=0,  }
where ${\hat E}=\det {\hat e_M}^A$. The fivebrane soliton appears
as a solution to the eleven-dimensional field equations and it
couples to the dual seven-form field strength $\hat F_7=*\hat F$.
Equation {\aixxiaa} then becomes the Bianchi identity for the
eleven-dimensional fivebrane.

This equation has in general gravitational Chern-Simons
corrections associated to the sigma-model anomaly on the
six-dimensional fivebrane worldvolume \ref\dlm{M.~J.~Duff,
J.~T.~Liu and R.~Minasian, ``Eleven Dimensional Origin of
String/String Duality: A One Loop Test'', \np {452} {1995}
{261}.}. The corrected fivebrane Bianchi identity takes the form
\eqn\aixxi{ d * \hat F=-{1\o 2} \hat F^2 +(2\pi)^4 { \b} {X_8}, }
where ${\beta}$ is related to the fivebrane tension by $T_6=1/
(2\pi)^3 {\beta }$. Henceforth we set ${\beta}=1$. Since the
gauge-fixed theory of the fivebrane is described by a chiral
anti-self-dual tensor multiplet, the eight-form anomaly polynomial
is expressed in terms of the Riemann tensor
\ref\agw{L.~Alvarez-Gaum\' e and E.~Witten,``Gravitational
Anomalies'', \np{234} {1983} {269}.}
\eqn\avii{ { X_8}={1\o (2\pi)^4}\left (-{1\o 768 } ({\rm tr} {\hat
R}^2 )^2 +{1\o 192} {\rm tr} {\hat R}^4\right).   }

The anomaly leads to an additional term in the action {\ai}
\eqn\aviii{ \d {\cal S}_{11} ={1\o 2} \int {\hat C} \wedge \left(
-{1\o 768} ({\rm tr}
 {\hat R}^2 )^2 +{1\o 192}
{\rm tr} {\hat R}^4 \right) . }
The existence of this interaction can be verified by computing the
one-point function of the two-form $B_{MN}$ in the Type IIA string
theory compactified on an eight-manifold \ref\vawi{C.~Vafa and
E.~Witten,``A One-Loop Test of String Duality'', \np{447} {1995}
{261}.}. The result of this calculation has no dilaton dependence,
since this would spoil gauge invariance. It can therefore be
extrapolated to eleven dimensions and it gives the previous
answer.

A supersymmetric configuration is one that obeys for some Majorana
spinor $\eta$ the conditions
\eqn\aix{ \eqalign{ & \d_{\eta} {\hat e_M}^A=0, \cr & \d_{\eta}
\hat C_{MNP}=0, \cr & \d_{\eta} \psi_M =0. \cr} }
Since in the background the spinor $\psi_M$ vanishes, the first
two of the above equations are satisfied, and only the gravitino
equation remains to be solved
\eqn\aaiixx{ {\hat \nabla} _M \eta - {1\o 288} \left( {{\hat
\G}_M}^{\,\,\,\,\, PQRS}-8 {\hat \d}_M^P {\hat \G}^{QRS} \right)
 \hat F_{PQRS} \eta =0.}
The most general ansatz for the metric, that is consistent with
maximal symmetry is
\eqn\gi{ {\hat g}_{MN}(x,y)=\D(y)^{-1} g_{MN} (x,y), }
where
\eqn\ax{g_{MN}(x,y) =\left( \matrix{ g_{\m \n}(x)  & 0 \cr 0 &
g_{mn}(y) \cr }\right).  }
Here $x$ are the three-dimensional external coordinates labeled by
the indices $\m,\n,\dots$ and $y$ the ones of the Euclidean
eight-manifold labeled by $m,n,\dots$. $\Delta(y)$ is a scalar
function called the ``warp factor''. We first would like to
rewrite {\aaiixx} in terms of $g_{MN}$. We can relate covariant
derivatives with respect to conformally transformed metrics by
using the formula
\eqn\di{{\hat \nabla }_M\eta =\nabla_M \eta +{1\o 2} \O^{-1}
{\G_M}^N (\nabla_N \O) \eta,  }
where ${\hat g}_{MN}=\O^2 g_{MN}$. This gives the relation
\eqn\dii{{\hat \nabla}_M \eta =\nabla_M \eta -{1\o 4} \D^{-1}
{\G_M}^N (\nabla_N \D) \eta. }
Furthermore, ${\hat \G}_M$ matrices are related to $\G_M$ matrices
as
\eqn\diii{{\hat \G}_M =\D^{-1/2} \G_M \qquad {\rm and} \qquad
{\hat \G}^M =\D^{1/2} \G^M, }
while ${\hat F}_{MNPQ}$ will be kept fixed under the
transformation {\gi}. We then obtain for {\aaiixx} in terms of
$g_{MN}$ the result
\eqn\div{ \nabla_M \eta -{1\o 4} {\G_M}^N \partial_N (\log \D
)\eta -{ 1\o 288} \D^{3/2}\left( {\G_M}^{PQRS} -8 \d_M^P \G^{QRS}
\right) F_{PQRS} \eta =0.}
We make a decomposition of the gamma matrices, that is appropriate
to the $11=3+8$ split, by taking
\eqn\axi{\eqalign{ &\G_{\m}=  \g_{\mu} \otimes \g_9,  \cr & \G_m=
1  \otimes \g_m,  \cr}}
where $\g_{\mu}$ and $\g_m$ are the gamma matrices of $M^3$ and
$K^8$ respectively and $\g_9$ is the eight-dimensional chirality
operator, that satisfies $\g_9^2=1$ and anti-commutes with all the
$\g_m$'s.

We decompose the eleven-dimensional spinor $\eta$ as a sum of
terms of the form
\eqn\axxii{\eta=\e \otimes \x,}
where $\e$ is a three-dimensional anticommuting spinor, while $\x$
is a commuting eight-dimensional Majorana-Weyl spinor. Spinors of
the form {\axxii}, that solve $\d_{\eta} \a=0$ for every field
$\a$, give unbroken supersymmetries. In this section we shall be
interested in compactifications having ${\cal N}=2$ supersymmetry
in three dimensions for which two spinors on $K^8$ of the same
chirality can be found. We can combine these real spinors into a
complex spinor of a well defined chirality. Without loss of
generality we will take the chirality to be positive.
Compactifications for which spinors of the previous form can be
found will, in general, have $\int X^8 \neq 0$.

In \ref\ipw{C.~J.~Isham and C.~N.~Pope, ``Nowhere-vanishing
Spinors and Topological Obstruction to the Equivalence of the NSR
and GS Superstrings'', \cqg {5} {1988} {257}; C.~J.~Isham,
C.~N.~Pope and N.~P.~Warner, ``Nowhere-vanishing Spinors and
Triality Rotations in 8-Manifolds'', \cqg{5} {1988} {1297}. } it
was shown, that demanding the existence of a nowhere-vanishing
eight-dimensional Majorana-Weyl spinor in the $8_c$ representation
of $SO(8)$ on a manifold with vanishing first Chern class gives a
relation between the Euler character $\chi$ of the eight-manifold
and the Pontryagin numbers, $p_1$ and $p_2$

\eqn\aax{p_1^2-4p_2+8 \chi =0. }
The Pontryagin numbers are obtained by integrating the first and
second Pontryagin forms {\agw}
\eqn\aaax{ P_1=-{1\o 2} {\rm tr} R^2\qquad {\rm and} \qquad
P_2=-{1\o 4} {\rm tr} R^4+{1\o 8} ({\rm tr} R^2)^2, }
over $K^8$. Replacing the spinor field in the $8_c$ representation
by a spinor in the $8_s$ representation of $SO(8)$ corresponds to
a change of sign in {\aax}
\eqn\aaxiix{p_1^2-4p_2-8 \chi =0.}
Therefore, if one asks for the existence of an $8_c$ and an $8_s$
nowhere-vanishing spinor field on the internal manifold, one
concludes, that the Euler character of $K^8$ has to vanish {\ipw}.
This observation will be relevant for section 3. However, it is
also true that for every manifold having $-8\chi=p_1^2-4p_2$ we
can find another one which has $8\chi=p_1^2-4p_2$, obtained by
reversing the orientation of the original manifold. This
corresponds to interchanging positive and negative chirality
spinors.

Comparing {\avii} with {\aaax} we observe that the anomaly
polynomial ${X}_8$ is proportional to $P_1^2-4P_2$ and is
therefore related to the Euler character of $K^8$
\eqn\aaix{ \int_{K^8} {X}_8=-{1\o 4! (2\pi)^4} \chi , }
which is a topological invariant. Finding nowhere-vanishing
spinors of both chiralities as a solution of {\div} thus implies,
that the integral of the anomaly polynomial {\aaix} vanishes and
will be used later on. Some examples of compactifications of
eleven-dimensional supergravity on eight-manifolds of this type
have been considered in \ref\pepo{C.~N.~Pope and
P.~van~Nieuwenhuizen, ``Compactifications of $d=11$ Supergravity
on K\" ahler Manifolds'', \cmp {122} {1989} {281}.}. For these
compactifications no warp factor has been taken into account and
the internal manifold is of the form $K^2\times K^6$, where $K^2$
is a two-dimensional sphere or torus and $K^6$ is a
six-dimensional Calabi-Yau manifold. They yield non-vanishing
expectation values for the four-form field strength, if the
external space is anti-de Sitter and have an ${\cal N}=4$
supersymmetry in three dimensions. As was shown in {\bb} and
\ref\katy{K.~Becker, ``A Note on Compactifications on Spin(7)
Holonomy Manifolds '', JHEP {\bf 0105} (003) 2001,
hep-th/0011114.} the situation is rather different, if the anomaly
is taken into account. In this case one can find solutions, that
preserve an ${\cal N}=2,1$ supersymmetry respectively, if the
external space is three-dimensional Minkowski space, while the
four-form field strength gets a non-vanishing expectation value.
Furthermore, we shall see in the next section, that if the
internal manifold is non-compact it is also possible to preserve
an ${\cal N}=1,2$ supersymmetry, while non-vanishing expectation
values for tensor fields are still present.

In compactifications with maximally symmetric three-dimensional
space-time the non-vanishing components of $F_4$ are
\eqn\bxii{\eqalign{ &F_{mnpq} \quad {\rm arbitrary,} \cr & F_{\mu
\nu \r m} =\e_{\m\n\r} f_m. \cr}}
In order to compare with the results of the next section, let us
briefly summarize the conditions imposed by ${\cal N}=2$
supersymmetry. First, some of the internal components of the flux
(and their complex conjugates) have to vanish $F_{1,3}=F_{4,0}=0$.
The only non-vanishing (internal) component of the flux is of type
$(2,2)$ and satisfies the primitivity condition
\eqn\bxiii{F\wedge J=0,}
where $J$ is the K\"ahler form. The internal manifold is not
K\"ahler but conformal to a Calabi-Yau four-fold with the warp
factor satisfying the equation
\eqn\biv{d*d \log \Delta={1\o 3} F\wedge F - {2\o 3} (2\pi)^4
X_8.}
The external component of the flux is related to the warp factor
\eqn\baiii{F_{\m \n \r m}=\e _{\m \n \r}\partial_m \D^{-3/2}. }
These are all the constraints imposed by ${\cal N}=2$
supersymmetry. Similar constraints for compactifications with
${\cal N}=1$ supersymmetry in three dimensions were derived in
{\katy}. Finally, to compare with the more generic situation of
the next section, let us remark, that in \ref\bbtwo{K.~Becker and
M.~Becker, ``Supersymmetry Breaking, ${\cal M}$-Theory and
Fluxes'', JHEP 0107:038, 2001, hep-th/0107044.} it was found, that
there are some solutions to the equations of motion, which break
supersymmetry and yet lead to a vanishing cosmological constant.
Any self-dual flux will be a solution to the equations of motion,
which means that its internal component is of the form
\eqn\bab{F=F_{4,0}+F_{0,4}+F_{2,2}+J\wedge J~F_{0,0},}
where $F_{2,2}$ is a primitive $(2,2)$ form. Only the second term
will preserve supersymmetry and yet all the components of the flux
above are allowed by the equations of motion. This situation will
be different in the next section, where we will see that we can
have anti-self dual fluxes on a Calabi-Yau four-fold(or via
orientation reversal, self-dual fluxes with right handed
covariantly constant spinors), where we still find that the flux
has to be of the (2,2) type, except now they will be of the form
$F = F_{(1,1)}\wedge J $.This is not in contradiction with
{\bbtwo} as one of the assumptions that went into showing that the
equations of motion require fluxes to be self-dual on a Calabi-Yau
four-fold(where the nowhere vanishing complex,chiral spinor
carried positive chirality), was that the manifold was compact,
which we shall show is not satisfied for our case.

\subsec{Non-Chiral Spinors on the Internal Space} Let us now
explore, how the previous results change, if we choose spinors on
the internal manifold of non definite chirality. Compactifying
again to three-dimensional Minkowski space the external component
of the gravitino supersymmetry transformation
\eqn\bxiiii{ \eqalign{ \d \psi_{\mu} ={\nabla}_{\m} \eta & - { 1\o
288}  \D^{3/2} (\g_{\m} \otimes \g_9 \g^{mnpq} ) F_{mnpq} \eta \cr
& +{ 1\o 6}\D^{3/2} ( \g_{\m} \otimes \g^m ) f_m \eta \cr & -{1\o
4}\partial_n (\log \D) ( \g_{\m} \otimes \g_9 \g^n ) \eta , \cr} }
leads to the following equation
\eqn\babb{F\x-48~{\g}^m(-{\g}^9f_m+{\partial}_m{\Delta}^{-{3/2}}){\x}=0.}
Here $F$ denotes the contraction of the internal flux component
with the antisymmetrized product of four gamma matrices
$F=F_{pqrs}{\g}^{pqrs}$ and $p,q,\dots$ denote real coordinates.
Notice, that here $F$ is not the four-form introduced in the
previous section. Nevertheless, it should be clear from the
context, what $F$ is refering to. In the previous formula we have
used a complex spinor $\x$ of indefinite chirality. We can
decompose $\x$ into its positive and negative chirality part
\eqn\babba{{\x}={\x}_++{\x}_-,}
by using the chirality projection operators
\eqn\baai{P_{\pm}{\x}={1\over 2}~(1~{\pm}{\g}^9)~{\x}.}
Acting with these operators on equation {\babb} we get the
following result for the positive and negative component of ${\x}$
\eqn\babii{F~{\x}_{\pm}={a \!\!\!\!\, /}_{\pm}~{\x}_{\mp}.}
 Here we have introduced the notation
\eqn\baaix{{a \!\!\!\!\, /}_{\pm}={\gamma}^m~a^{\pm}_m=
48~{\gamma}^m~({\pm}f_m+{{\partial}_m}{\Delta}^{-3/2}).}
We can reduce these equations to determining equations for the
self-dual and anti-self-dual part of $F$
\eqn\baxx{G^{\pm}_{pqrs}=F_{pqrs}{\pm}*F_{pqrs},}
which take the form
\eqn\baxxi{G^{\pm}_p~{\x}_{\pm}=
G^{\pm}_{pqrs}~{\g}^{qrs}~{\x}_{\pm}={\g}_p~{1\over 4}~{a
\!\!\!\!\, /}_{\pm}~{\x}_{\mp}.}
The upper sign determines the self-dual part of $F$, while the
lower sign determines its anti-self-dual part. Thus, these are two
of the general equations, that need to be solved to determine the
supersymmetry constraints. Equivalently, we can rewrite these
equations in a way, that will be useful in order to simplify the
internal component of the gravitino transformation
\eqn\baxxi{F_n~{\x}_{\pm}={\g}_n~{1\over 8}~{a \!\!\!\!\,
/}_{\pm}~{\x}_{\mp}+{1\over 2}G^{\mp}_n~{\x}_{\pm}.}
Let us now have a look at the other two equations, which come from
the internal component of the gravitino supersymmetry
transformation. Using some gamma matrix algebra these can be
written as
\eqn\baxxia{{\nabla}_m{\tilde \x}+{1\over
24}{\Delta}^{3/2}F_m{\tilde \x}+{1\over
4}{\Delta}^{3/2}f_n~{\g}_m^n{\gamma}_9~{\tilde \x}=0,}
where $F_m=F_{mpqr}{\g}^{pqr}$ and we have rescaled the spinor
${\x}={\Delta}^{-1/4}{\tilde \x}$. Writing this in terms of
negative and positive chirality spinors we obtain
\eqn\bbbs{{\nabla}_m{\tilde \x}_{\pm}+{1\over
24}{\Delta}^{3/2}F_m{\tilde \x}_{\mp}~{\pm}{1\over
4}{\Delta}^{3/2}f_n~{\g}_m^n~{\tilde \x}_{\pm}=0.}
After rescaling the internal metric $\tilde
g_{mn}={\Delta}^{-3/2}g_{mn}$ and using equation {\baxxi} we can
write the final expression for the equations coming from the
internal component of the gravitino as
\eqn\bbbsi{{\tilde {\nabla}}_m{\tilde \x}_{\pm}+{1\over
192}{\Delta}^{3/2}a^{\mp}_m{\tilde \x}_{\pm}+{1\over
48}{\Delta}^{-3/4}G^{\pm}_m~{\tilde \x}_{\mp}=0.}
So, to summarize, this equation together with {\babii} in terms of
the rescaled metric
\eqn\bcii{{\tilde F}{\tilde \x}_{\pm}={\tilde
\gamma}^{mnpq}F_{mnpq}~{\tilde \x}_{\pm}={\Delta}^{9/4} {\tilde {a
\!\!\!\!\, /}_{\pm}}{\x}_{\mp},}
need to be solved to find the constraints on the fluxes and the
internal manifold. From now on we shall consider the case, where
the internal fluxes are self-dual

\eqn\baxx{G^{-}_{pqrs}=0 .}
In this case there is a relation between the external component of
the flux and the warp factor
\eqn\baxxix{f_m={\partial}_m{\Delta}^{-3/2},}
which leads to ${a \!\!\!\!\, /}_-=0$ and
\eqn\baxxib{{a \!\!\!\!\,
/}_+=96~{\g}^m{\partial}_m{\Delta}^{-3/2}.}
For convenience we will rescale the spinor $\x_-$ again
\eqn\bix{\x'_-={\Delta}^{-3/4}\x_-,}
as the equations then simplify further. For these self-dual
solutions there are three equations, that need to be solved
\eqn\bixa{F_n~\x_++{1\over 8}~{\gamma}_n~{a \!\!\!\!\, /}_+
~\x_-=0,}
which comes from the external component of the gravitino, while
the internal component gives two further equations
\eqn\bbbsi{{\nabla}_m{\x}_++{1\over 24}F_m~{\x_-}=0,}
and
\eqn\bbbsii{{\nabla}_m{\x}_-=0.}
This expression implies, that the spinor ${\x}_-$ is covariantly
constant. If ${\xi_-}$ happens to vanish everywhere, then we get
back the flux conditions of {\bb}, by analysing equations {\bbbsi}
and {\bixa}.Since we are looking to generalise the analysis in
{\bb}, we shall therefore take ${\xi_-}$ to be nowhere vanishing,
which is the only other possibility. Furthermore, we have dropped
the tildes and primes from the spinors and gamma matrices in order
to simplify the notation. From the previous two equations we make
two observations. First, from  {\bbbsii} we see, that we have
found a spinor, $\x_-$, that is covariantly constant. If this
spinor is complex we are dealing with a conformally Calabi-Yau
manifold, while if the imaginary part of $\x_-$ vanishes, it is
possible to derive from the above equations a generalization of
the supersymmetry constraints for a Spin(7) holonomy manifold
computed in {\katy}. We will elaborate on this a little bit more
in a moment. Second, if the spinors $\x_+$ and $\x_-$ are both
nowhere vanishing it implies according to {\ipw}, that the Euler
characteristic of the internal manifold is zero. Thus, in order to
obtain non-trivial solutions, where fluxes are turned on, we need
to consider non-compact internal manifolds or we need to take
sources into account. However, in general, +${\x}_+$ can have
zeros, so that the Euler character of the internal manifold does
not vanish.

Due to the existence of these two spinors we can define (for an
appropriate normalization of the spinors) a vector field on our
internal manifold as follows
\eqn\caxvi{\upsilon_n={\x_+}^{\dagger} {\gamma_n}~{\x_-},}
where $n=3,\dots, 10$ runs over the internal coordinates. Here we
have combined the two spinors on our manifold (we are dealing with
${\cal N}=2$ supersymmetry) into a complex spinor, whose chiral
parts we called ${\x}_+$ and ${\x}_-$ as before. In case, that the
covariantly constant spinor is real, we have a manifold, that is
conformal to a Spin(7) holonomy manifold. We will work out the
generalization of the sypersymmetry constraints derived in {\katy}
at the end of this section. In writing the vector field as above,
we have chosen a normalization of $1/8$ for the spinor ${\x}_-$.
Fierz identities imply, that the positive chirality spinor can be
written in terms of this vector field
\eqn\caxvib{\x_+=\upsilon^n{\gamma_n}~{\x_-}.}
Using this expression for the positive chirality spinor and
{\bixa}, {\bbbsi} and {\bbbsii} we obtain the following two
equations, which are defined only in terms of the negative
chirality spinor ${\x_-}$
\eqn\dia{[{\gamma}_n({\nabla}_m \upsilon^n)+{1\over
24}F_m]~{\x_-}=0,}
and
\eqn\bixai{[F_m\upsilon^n{\gamma}_n+{1\over 8}~{\gamma}_m]~{a
\!\!\!\!\, /}_+ ~\x_-=0.}
We can now introduce complex coordinates and take into account,
that ${\x}_-$ is annihilated by ${\gamma}_{\bar a}$
\eqn\diia{{\gamma}_{\bar a}~\x_-={\gamma}^a{\x}_- = 0.}
>From equation {\di} we get several constraints on the fluxes.
First, the $(2,2)$ part of the flux is no longer primitive but
satisfies the condition
\eqn\diix{{\nabla}_a\upsilon_{\bar b}+{1\over 8}~F_{a{\bar
b}c{\bar d}}~J^{c{\bar d}}=0.}
This can be written equivalently as
\eqn\eiv{F\wedge J+*dv=0,}
where we have rescaled $v$ with a factor $15/8$. This equation is
a determining equation for the fundamental form $J$, once the
vector field is specified (Note that equation {\diix}implies that
the vector field is Killing, since equation {\diix} tells us that
the symmetric part ${\nabla_{(a} {\upsilon}_{{\bar b})}}$ as well
as the part ${\nabla_{a} {\upsilon}_{b}}$ both vanish, which is
equivalent to the Killing constraint.In fact this defines a
holomorphic Killing vector, though we will not make use of it
here.There is another constraint on ${\upsilon}$ coming from the
external component of the gravitino equation,which we'll examine
below) and the fluxes are given. It is a generalization of the
primitivity condition {\bxiii} obeyed by the fluxes for the case,
that the spinors on the internal space are no longer chiral.

Furthermore, it is easy to see, that as in the case of chiral
spinors, the $(4,0)$ and $(3,1)$ parts of the flux (and its
complex conjugate) have to vanish
\eqn\diixa{F_{(4,0)}=F_{(3,1)}=0,}
while the derivative of the vector field with respect to the
complex conjugate coordinate vanishes
\eqn\diix{{\nabla}_{\bar a}\upsilon_{\bar b}=0.}
Finally, we can derive the determining equation of the vector
field by using {\bixa}. This leads to the following equation
\eqn\dix{(F_{mnpq}{\upsilon_r} {\gamma^{npq}} {\gamma^r} + 12
{\gamma_m}{\gamma_n}{\partial^n}{\Delta^{-3/2}} ){\xi_-} = 0.}
Using formula {\diix}, some gamma matrix algebra and taking into
account, that the gamma matrices act as raising and lowering
operators, allows us to write the above expression in the
following form
\eqn\dixx{ {\partial_{m}{\Delta^{-3/2}}} -8
{\upsilon}^{n}{\partial_{[m}{\upsilon_{n]}}} = 0.}
Equation {\dixx} determines the vector field ${\upsilon_{m}}$, as
the warp factor can be computed in terms of the internal component
of the flux through equation {\biv}.

If we reversed the orientation of our manifold, then the above
equations would be left unchanged, except that, now the flux would
be required to be anti-self dual, while the covariantly constant
spinor would have positive(left) chirality.Then, the content of
our equations would be that the $(4,0)$ and $(3,1)$ part of the
flux vanish, while the $(2,2)$ part of the flux is non-primitive,
and being anti-self dual, must be of the form $F_{(1,1)}\wedge J$
as claimed before.This is the picture we'll have in mind when we
refer to the flux as being non-primitive.That is, we will talk
about our internal manifold, carrying anti self-dual fluxes, with
the covariantly constant spinors being of positive(left)
chirality, as is conventional.

In order that the equation {\eiv} properly generalise the solution
of {\bb}, it must be that, when the internal manifold is compact,
this solution must go back to the flux condition in {\bb}. This is
because, for a compact four-fold, the superpotential, and hence
the equations of motion can be deduced from a dimensional
reduction of the 11D action {\ai} as was done in \ref\hl{M.~Haack
and J.~Louis,``M-theory compactified on Calabi-Yau fourfolds with
background flux'', Phys.\ Lett.\ B {\bf 507}, 296 (2001),
hep-th/0103068.}.That the internal manifold cannot be compact can
be seen as follows.Firstly, we have a Killing vector on a
ricci-flat manifold.If the manifold also happens to be compact,
then such a vector is also covariantly constant.Secondly, a
manifold whose holonomy is $SU(4)$, and not a proper subgroup of
it, does not admit covariantly constant vectors(since under the
holonomy group, $SO(8)$ decomposes as $ 8 =  4 + {\bar 4}$ , which
does not contain a singlet).For more details on this proof, refer
to \ref\gsw{M.~B.~Green, J.~H.~Schwarz and E.~Witten,
``Superstring Theory. Vol. 2: Loop Amplitudes, Anomalies And
Phenomenology''}.Another way to see this same result is to note
that, when the manifold is compact, the equations of motion
analysed in {\bbtwo} imply that the internal flux has to be
self-dual(with the orientation determined by the holomorphic
4-form).However, our starting point was the assumption that the
flux was anti-self dual, and hence this implies the internal flux
vanishes, leading to constant warp factor.

To summarize: Assuming that the fluxes on our internal manifold
are (anti)self-dual, we have shown, that the internal manifold is
conformal to a Calabi-Yau four-fold {\bbbsii}, with two spinors of
non-definite chirality, whose (left)right chiral parts are
covariantly constant. The only non-vanishing component of the flux
is the $(2,2)$ component, which is no longer primitive {\eiv},
{\diixa} but still preserves supersymmetry. Equation {\dixx}
determines the vector field, while the warp factor is determined
by {\biv}.

Let us now derive the generalization of the supersymmetry
constraints for the Spin(7) holonomy case considered in {\katy}.
First, from {\dia} we have a
\eqn\dia{[{\gamma}_n({\nabla}_m \epsilon^n)+{1\over
24}F_m]~{\x_-}=0.}
Multiplying {\dia} with ${\x_-}^T~{\gamma}^a$ we get
\eqn\dixxaa{{\nabla}_k \epsilon^m + {1\over
24}F_{knpq}{\Omega}^{mnpq}=0,}
where
${\Omega}^{mnpq}={\x_-}^T{\gamma}^{mnpq}{\x}_-/{{\x^T_-}{\x}_-}$
is the self-dual four-form of the internal space. Substituting for
${\nabla}_m {\epsilon_n}$ in {\dia} using {\dixxaa}, and
multiplying the resulting expression by ${\gamma}_m$ and observing
that the flux is anti self-dual while the real four form
${\Omega}$ is self-dual, we get:
\eqn\dixxb{ {F_{pqr[m}{\Omega^{pqr}_{n]}}} = 0.}
\eqn\dixxc{ {\nabla}_{[m}{\epsilon_{n]}} = 0 .}
>From {\dixxc} and {\dixxaa},we can see that the other constraint
can be cast into the following equation:
\eqn\dixxd{{\nabla}_{(k} \epsilon_{m)} + {1\over
24}{F_{npq}}_{(k}{\Omega}_{m)}^{npq}=0,}
In order to recover the equations derived in {\katy}, we need to
go back to equation {\bbbsii} and set ${\xi_-} = 0$.In equation
{\bbbsii}, we had a choice of setting ${\xi_-}$ zero everywhere,
or nowhere zero.It is the former choice that gives us the flux
conditions of {\bb} and it is straightforward to check that it
does.If we set ${\upsilon} = 0$ in our equations, we would not get
back {\bb}.Instead, we should end up with zero flux and constant
warp factor, as can be easily checked, both in equations {\eiv}
and {\dixx}, as well as in equations {\dixxb},and {\dixxd}.To be
more precise, note that {\eiv} and {\diixa} together imply that
the flux is primitive and of type (2,2), when ${\upsilon}$
vanishes.On a Calabi-Yau four-fold, the primitive (2,2) forms are
self-dual, this means our flux has to be self-dual to solve the
equations for supersymmetry.However, by assumption, our fluxes
were anti self-dual, which means that in order to preserve
supersymmetry the flux has to vanish.

Another point to note, is that the conditions given in equations
{\dixxb},{\dixxc} and {\dixxaa} must also be solved by the ansatz
we had in the previous section, when the manifold was a Calabi-Yau
four-fold.It is easy to check that this is indeed the case, once
we notice that if the vector field ${\epsilon}$ is given by
${\epsilon_m} = i({\upsilon_m} - {\bar{\upsilon}_m})$, then all
the relevant equations are automatically satisfied.This is because
among the content of the $N=2$ supersymmetry constraints was the
fact that the vector field ${\upsilon}$ was real, which means
${\epsilon}$ vanishes.We in fact used this in writing {\diix} in
the form of {\eiv}.

Finally, let us remark that the form of the fluxes which solve the
equations of motion are no longer of the form {\bab}.This is
because we are no longer dealing with a compact manifold, and the
assumptions that went into showing that the fluxes has to be
self-dual on a Calabi-Yau four-fold {\bbtwo} are not satisfied for
our case, allowing us to have anti self-dual fluxes on a
Calabi-Yau four-fold and still solve the equations of
motion.However, the presence of a flux of (3,1) type breaks
supersymmetry while a (2,2) flux of the form $F_{(1,1)} \wedge J$
preserves supersymmetry.

\newsec{Example: PP-Waves and ${\cal M}$-Theory} By restricting
ourselves to solutions for which the internal fluxes are
self-dual, we have managed to make the problem of solving the
gravitino equations tractable. In this case we end up with
manifolds with special holonomy as in {\bb} but as we can see from
the above derivation the flux conditions obeyed by our models are
different from the ones considered in {\bb}. In particular, the
internal fluxes are constrained to be of the $(2,2)$ form, but no
longer have to be primitive. We now would like to consider a
particular example of this more general class of models.

More concretely, we would like to consider an example constructed
in {\cv} where the internal manifold is flat $R^8$, thus retaining
maximal supersymmetry. This solution arises as the ${\cal
M}$-theory lift (by performing a timelike T-duality) of the
supersymmetric PP-wave solution, which comes from the Penrose
limit of the ${AdS_5}{\times}{S_5}$ compactification of the Type
IIB theory. As noticed in {\cv} this model does not obey the flux
constraints derived in {\bb} but we shall see, that the
supersymmetry constraints derived in the previous section are
satisfied. The reason for this is, of course, that the spinors on
the internal manifold used in {\cv} are not chiral. In the
following we will use the notation and conventions of {\cv}.

The lift of the Type IIB PP-wave solution to ${\cal M}$-theory has
a metric of the form of a deformed M2-brane
\eqn\caxi{{{ds_{11}^2}}  = H^{-2/3} (-dt^2 + d{x_1}^2 + d{x_2}^2 )
+ H^{1/3} \Sigma_i d{ z_i}^2,}
where H is the warp factor and the coordinates, ${z_i}$ with
$i=1,\dots, 8$ run over the internal manifold, which is flat and
$t,x_1, x_2$ describe the three-dimensional external space. The
fluxes are of the form
\eqn\caxii{F = dt {\wedge} dx_1 {\wedge} dx_2 {\wedge} dH^{-1} +
{\mu} {\Phi}_{(4)},}
where ${\Phi}_{(4)} = \Phi_{mnpq}$ is the self-dual flux on $R^8$,
which is assumed to be constant in order to preserve all the
supersymmetry
\eqn\gix{{\Phi}_{(4)}=dz^1\wedge dz^2 \wedge dz^3 \wedge dz^4+
dz^5\wedge dz^6 \wedge dz^7 \wedge dz^8,}
and $\m$ is a constant. The warp factor satisfies the equation of
motion
\eqn\caxiii{{\Delta} H  =  -{1\o 48 } { \mu }^2 {\Phi}_{(4)} ^2.}
Here one has to take into account, that in this example two
no-where vanishing spinors can be found (as we shall see below),
so that according to {\ipw}, the Euler characteristic of the
internal manifold is zero. Thus, in order to obtain non-trivial
solutions, where fluxes are turned on, one needs to consider
non-compact internal manifolds, in this case flat $R^8$.

As shown by {\cv}, the gravitino supersymmetry equations are
satisfied by making the following ansatz for the internal spinor
\eqn\caxiv{ {\epsilon} = H^{-1/6} ({\nabla_i}H W {\gamma^i} -
H^{1/2} ) {\eta},}
where $ W = {\Phi}_{ijkm}{\gamma}^{ijkm} $ and ${\eta}$ is a
constant spinor of negative chirality. From the above equations we
can determine the vector field, that arises in our equation
{\caxvib}. To do so, notice the expressions for the spinors are
${\xi_-} = {\eta} $ and ${\xi_+} =
{W{\gamma^i{\nabla_i}H{\eta}}\over {4\m}}$.
This leads  to the following equation for the vector field
\eqn\caxvi{\upsilon_{i} = {{\nabla}^{j}H {\eta}^{\dag}
{\gamma_{i}}W{\gamma_{j}}{\eta} \o {4\mu}} .}
Differentiating the above expression, antisymmetrizing and noting,
that $F$ is related to ${\Phi}_{(4)}$ by the proportionality
constant ${\mu}$, gives us the condition found in the last
section, that replaces primitivity
\eqn\caxvii{*d{\upsilon} + F {\wedge } J = 0.}
In the last step we have absorbed a factor of $3/4$ into the
definition of ${\upsilon}$. It can also be seen, that the
determining equation for the vector field is satisfied. The
easiest way to check this is to substitute the expression for
${\x}_+$ into {\bixa} and taking into account, that
$H={\Delta}^{3/2}$ in our notation.Finally, we must point out that
it is possible to choose complex coordinates in the above example
in terms of which, the internal flux is a $(2,2)$ form, and this
also allows us to show that the equations for ${\cal N}= 2$
supersymmetry are satisfied.

So we conclude, that the deformed M2-brane constructed in {\cv}
arises as a special case of our general set of models, by setting
the internal manifold to be flat. Using the constraints, that we
derived in the previous sections, more examples of models
satisfying our equations could be constructed.The above example
was one that retained a lot of supersymmetry, and it would be
interesting to know if there do exist solutions with precisely
${\cal N} = 2$ supersymmetry, that satisfy our constraints.It is
clear that such solutions will have to describe non-compact
internal manifolds.

\newsec{Discussion and Conclusions}
Compactifications with non-vanishing fluxes are a rather
fascinating area of recent research, as they allow us to approach
and even solve one longstanding problem in string theory, the
moduli space problem. The simplest type of such models correspond
to ${\cal M}$-theory compactifications on non-K\"ahler conformally
Calabi-Yau manifolds, that have four complex dimensions. In this
case two covariantly constant Majorana-Weyl spinors on the
internal space can be found. The constraints obeyed by the fluxes
and the geometry were derived in {\bb} and {\svw}. Similarly,
having only one covariantly constant spinor of definite chirality
leads to the supersymmetry constraints obeyed by manifolds, that
are conformal to Spin(7) holonomy manifolds {\katy}.See also
\ref\bhs{M.~Berg, M.~Haack and H.~Samtleben,``Calabi-Yau fourfolds
with flux and supersymmetry breaking'', JHEP {\bf 0304}, 046
(2003), hep-th/0212255},where it is shown that fluxes of the form
$(4,0)$ and $(0,0)$ can break ${\cal N} = 2$ supersymmetry into
${\cal N} = 1$.

In this note we have relaxed the assumption regarding the
chirality of the spinors on the internal manifold for
compactification of ${\cal M}$-theory to three-dimensional
Minkowski space-time with ${\cal N}=1,2$ supersymmetry and have
solved the supersymmetry constraints for a special class of
manifolds, that arise, when we drop the chirality assumption, yet
retain the assumption, that the internal fluxes are self-dual. The
resulting class of manifolds includes constructions like the
models {\cv}, where the Type IIB $PP$-wave is lifted to ${\cal
M}$-theory as a deformed membrane. If we remove the self-duality
assumption on the fluxes we expect to get the most general class
of ${\cal M}$-theory compactifications to three flat dimensions.
We expect, that the dual description of the Polchinski-Strassler
model would lie in this class of solutions. This would be helpful
in order to find the complete form of the Polchinski-Strassler
model, as to this date only the first orders in perturbation
theory are known. Approaching this problem from the Type IIB side,
where the work of \ref\rds{K.~Dasgupta, G.~Rajesh and S.~Sethi,
``M-theory, Orientifolds and G-Flux'', JHEP {\bf 9908} (023) 1999,
hep-th/9908088.} needs to be generalized by allowing spinors of
non-definite chirality on the internal space, might be the easiest
way to proceed. The recent paper of {\fg} should be rather
interesting for this purpose, as they have obtained the most
general ${\cal M}$-theory compactifications preserving ${\cal N}
=2$ supersymmetry.In the case of more general compactifications,
it is no longer clear, that the internal manifold is even a
complex manifold. Also, if we relax the assumption of
self-duality, then we do not expect to end up with manifolds of
special holonomy. In that case we would have to seek a
classification of the manifold in terms of $G$-structures.A
general discussion of G-structures, and their relevance to Type
IIB compactifications with NS-fluxes,is provided in
\ref\gmpw{J.~P.~Gauntlett, D.~Martelli, S.~Pakis and
D.~Waldram,``G-structures and wrapped NS5-branes''
,hep-th/0205050} .Classification of the most general
supersymmetric geometries from the ${\cal M}$-theory side was
performed in \ref\gp{J.~P.~Gauntlett and S.~Pakis,``The geometry
of D = 11
 Killing spinors'',JHEP {\bf 0304}, 039 (2003),hep-th/0212008.}
using G-structures.

\vskip 1cm

\noindent {\bf Acknowledgement}

\noindent

We thank K.~Dasgupta, S.~Kachru, A.~Krause, J.~Polchinski and
M.~Strassler for useful discussions on related subjects. The work
of K.B. was supported by the NSF under grant PHY-0244722, an
Alfred Sloan Fellowship and the University of Utah. The work of
M.B and R.S. was supported by NSF under grant PHY-01-5-23911, an
Alfred Sloan Fellowship and the University of Maryland.

\vskip 1cm

{\noindent} {\bf Appendix}

\noindent In this appendix we would like to collect a few useful
formulas and we would like to explain our notation.
\item{$\triangleright$} The different types of indices that we use
are \item{} $M,N,\dots$ are eleven-dimensional indices \item{}
$m,n,\dots$ denote eight-dimensional indices \item{}
$\mu,\nu,\dots$ are the indices of the external space
\item{$\triangleright$} $n$-forms are defined with a factor
$1/n!$.

For example

$$
F={ 1\over 4!} F_{mnpq} dx^m \wedge dx^n \wedge dx^p \wedge dx^q.
$$

\item{$\triangleright$} The gamma matrices $\Gamma_M$ are
hermitian while $\Gamma_0$ is antihermitian. They satisfy

$$
\{ \Gamma_M, \Gamma_N \} = 2 g_{MN}.
$$
\item{$\triangleright$} $\Gamma_{M_1 \dots  M_n}$ is the
antisymmetrized product of gamma matrices
$$
\Gamma_{M_1 \dots  M_n} =\Gamma_{[ M_1 \dots  } \Gamma_{M_n]}
$$
where the square bracket implies a sum over $n!$ terms with a
$1/n!$ prefactor. \item{$\triangleright$}Gamma matrix identities
that are useful are
$$
\eqalign{ & [ \gamma_m, \gamma^r]=2 {\gamma_m}^r \cr & [
\gamma_{mnp}, \gamma^{rs} ]=12 {\delta_{[m}}^{[r}
{\gamma_{np]}}^{s]}\cr & \{ \gamma_{mnpq}, \gamma^{rst} \} = 2
{\gamma_{mnpq}}^{rst} -72{\delta_{[mn}}^{[rs} {\gamma_{pq]}}^{t]}}
$$
\item{$\triangleright$} Our definition of Hodge $\star$ in $d$
dimensions is
$$
\star ( dx^{m_1} \wedge \dots \wedge dx^{m_p} ) = {| g|^{1/2}\over
(d-p)!} {{\epsilon^{m_1 \dots m_p}}_{m_{p+1} \dots m_d}
dx^{m_{p+1} }  \wedge \dots \wedge dx^{m_d} , }
$$
where
$$
\epsilon_{m_1 m_2 \dots m_d}= \cases{0 & any two indices repeated
\cr +1 &  even permutation \cr -1 & odd permutation \cr }
$$

\item{$\triangleright$} The identity which relates covariant
derivatives of spinors with respect to conformally transformed
metrics is

$$
\eqalign{ & {\tilde \nabla}_M \epsilon = \nabla_M \epsilon + {1
\over 2} \Omega^{-1} {\Gamma_M}^N ( \nabla_N \Omega) \epsilon ,
\cr & {\tilde g}_{MN} =\Omega^2 g_{MN} . \cr}
$$

\listrefs

\end